# Efficient energy propagation through self-assembled gold nanoparticle chain waveguides

Fatih N. Gür[1], Cillian P. T. McPolin[2] Søren Raza[3,‡], Martin Mayer[1,4], Diane J. Roth[2], Anja Maria Steiner[1,4], Markus Löffler[1], Andreas Fery[1,4,5], Mark L. Brongersma[3], Anatoly V. Zayats[2], Tobias A. F. König[1,4] & Thorsten L. Schmidt[1,6]

[1]*Center for Advancing Electronics Dresden (cfaed), Technische Universität Dresden, 01062 Dresden, Germany*

[2]*Department of Physics, King's College London, Strand, London, WC2R 2LS, UK*

[3]*Geballe Laboratory for Advanced Materials, Stanford University, Stanford, CA 94305-4045*

[4]*Leibniz-Institut für Polymerforschung Dresden e.V., Institute of Physical Chemistry and Polymer Physics, Hohe Str. 6, 01069 Dresden, Germany*

[5]*Department of Physical Chemistry of Polymeric Materials, Technische Universität Dresden, Hohe Str. 6, 01069 Dresden, Germany*

[6]*B CUBE – Center for Molecular Bioengineering, Technische Universität Dresden, 01062 Dresden, Germany.*

‡Present address: Department of Micro- and Nanotechnology, Technical University of Denmark, DK-2800 Kongens Lyngby, Denmark.

**Abstract:**

**The strong interaction of light with metallic nanoparticles enables field confinement well below the diffraction limit. Plasmonic waveguides consisting of metal nanoparticle chains could be used for the propagation of energy or information on the nanoscale, but high losses have thus far impeded practical applications. Here we demonstrate that efficient waveguiding is possible through gold nanoparticle chains despite the high dissipative losses of gold. A DNA origami directed self-assembly of monocrystalline, spherical nanoparticles allows the interparticle spacing to be decreased to 2 nm or below, which gives rise to lower-energy plasmon resonance modes. Our simulations imply that these lower energy modes allow efficient waveguiding but collapse if interparticle gap sizes are increased. Individual waveguides are characterized with nanometer-resolution by electron energy loss spectroscopy, and directed propagation of energy towards a fluorescent nanodiamond and nanoscale energy conversion is shown by cathodoluminescence imaging spectroscopy on a single-device level. With this approach, micrometer-long propagation lengths might be achieved, enabling applications in information technology, sensing and quantum optics.**

Light can transmit information much faster and more energy efficiently through glass fibres compared to electric data transmission. Diffraction, however, impedes the miniaturization of optical components to the nanometre scale and thus the integration into nanometre-sized electronic components. Plasmonic structures are frequently discussed as a potential strategy for the miniaturization of optical or optoelectronic components[1–4]. The field of plasmonics exploits the interaction of light with nanoscale metallic structures to confine, guide and manipulate light on scales below the diffraction limit[5,6], thereby greatly benefiting applications such as quantum photonics, or short-distance optical communication[1–4]. For example, deep subwavelength plasmonic waveguides fabricated from closely spaced metal nanoparticle chains have been proposed two decades ago,[7] and realized mainly by electron beam lithography[6,8–10]. Lithography is, however, a slow, expensive, and non-scalable method. Moreover, the metal nanoparticle shape cannot be precisely controlled by the lift-off procedures and the interparticle gaps are usually limited to 10 nm or more.

Chemical synthesis of colloidal nanoparticles offers a more precise control over the metal particle crystallinity, shape, and size. Such colloidal nanoparticles can be self-assembled by leveraging the programmability and scalability of DNA nanotechnology[11–21]. Often, the robust DNA origami method is used to pattern a large variety of inorganic nanoparticles or other functional elements including proteins, or small molecules, with nanometre precision on two or three dimensions[13,22]. In this manner, plasmonic devices with a static[16] or reconfigurable[23] circular dichroism; plasmonic hotspots for the sensing of fluorescence molecules[15,21] were created from metal nanoparticles; and a coherent energy oscillation between two gold nanoparticles (AuNPs) was demonstrated through one 30 nm silver nanoparticle[20]. Recently, self-assembled gold nanoparticle chain waveguides on DNA origami were reported, however, without showing energy propagation[17], or only showing a minimal energy transfer over 35-50 nm[19].

Thus far, experimentally realized[8–10,6,17,19] and theoretically simulated[8,24] nanometre-sized gold particle chain waveguides show a high energy loss that reduces the incident field after propagation over just 50 nm by around 1-2 orders of magnitude, which impeded practical applications. Here, we demonstrate that losses can be reduced to a few percent per 50 nm. For this, we self-assembled plasmonic waveguides on six-helix bundle (6-HB) DNA origami nanotubes, which display binding sites for eight oligonucleotide-functionalized gold nanoparticles (AuNPs; details in Figure S 1 and Figure S 2)[18], and studied their optical properties at a single-device level with nanometre resolution. As the plasmon coupling between nanoparticles increases with decreasing inter-particle distance[16], we minimized the inter-particle spacing by using AuNPs with a diameter equal to the distance between binding sites. We prepared both complete waveguides, with eight AuNPs, and incomplete waveguides where we omitted the binding sites for two particles in the centre of the waveguide, serving as a negative control (for details, see figs. S1-S2). As the combined diameter of the AuNPs and the oligonucleotide shell is larger than the distance between the binding sites on the 6-HB, the particles are assumed to minimize their free energy during the drying process,[25] thereby minimizing the inter-particle spacing, which occasionally results in slightly deformed zigzag patterns (Fig. 1). This mechanical compression of the protective oligonucleotide layer around the AuNPs allows us for the reduction of the gap sizes from around 7 nm, for mechanically uncompressed functionalized AuNPs,[26] to 2 nm or less. Such a small inter-particle spacing cannot be routinely achieved by conventional lithography methods.

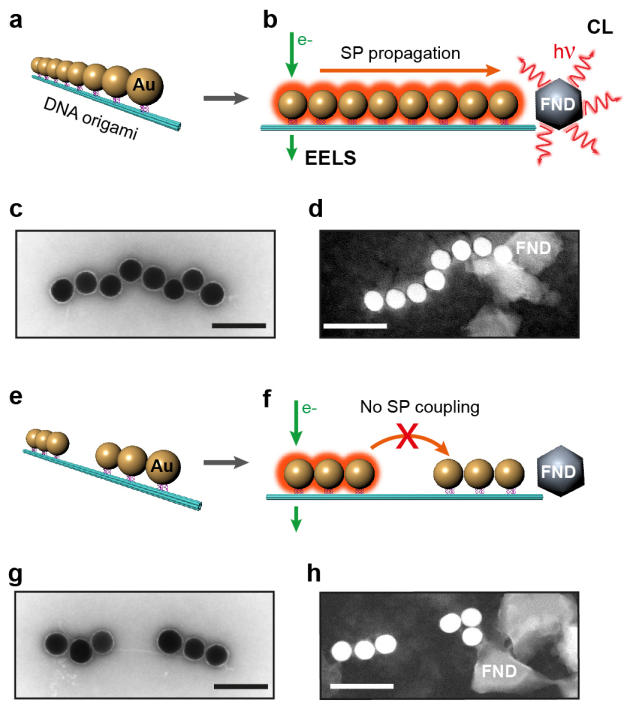

**Figure 1 | Overview of self-assembled plasmonic waveguides.** Schematic representation and corresponding transmission mode scanning electron microscope (tSEM) bright-field micrographs and tSEM dark-field micrographs of complete **(a-d)** and incomplete **(e-h)** waveguides. Surface plasmon (SP) resonance modes are excited by an electron beam and the loss of energy is detected by electron energy loss spectroscopy (EELS). In addition, the SPs propagating towards a fluorescent nanodiamond (FND) excite a fluorescence signal that is detected by cathodoluminescence (CL) imaging spectroscopy. SPs do not couple efficiently across the large gap, resulting in the absence of signal from the FND when the isolated sections of the incomplete waveguides are illuminated. Scale bars, 100 nm.

In contrast to any optical technique, electron energy loss spectroscopy (EELS)[27,28] provides the simultaneous high spatial and spectral resolution needed to accurately map the relevant surface plasmon (SP) modes of the waveguide[29] at a single nanoparticle level. To detect energy transport along the waveguide *via* the SP modes, we place a fluorescent nanodiamond (FND)[30] at the end of the waveguide. After excitation of the waveguide, the FND decays by emitting photons into the far-field, which is measured by cathodoluminescence (CL) imaging spectroscopy[31] (Fig. 1). For EELS measurements, the waveguides were deposited on a thin silicon nitride membrane and measured with a transmission electron microscope (TEM). We observed different SP modes and their excitation depends strongly on the position of the electron beam (see Figure 2).

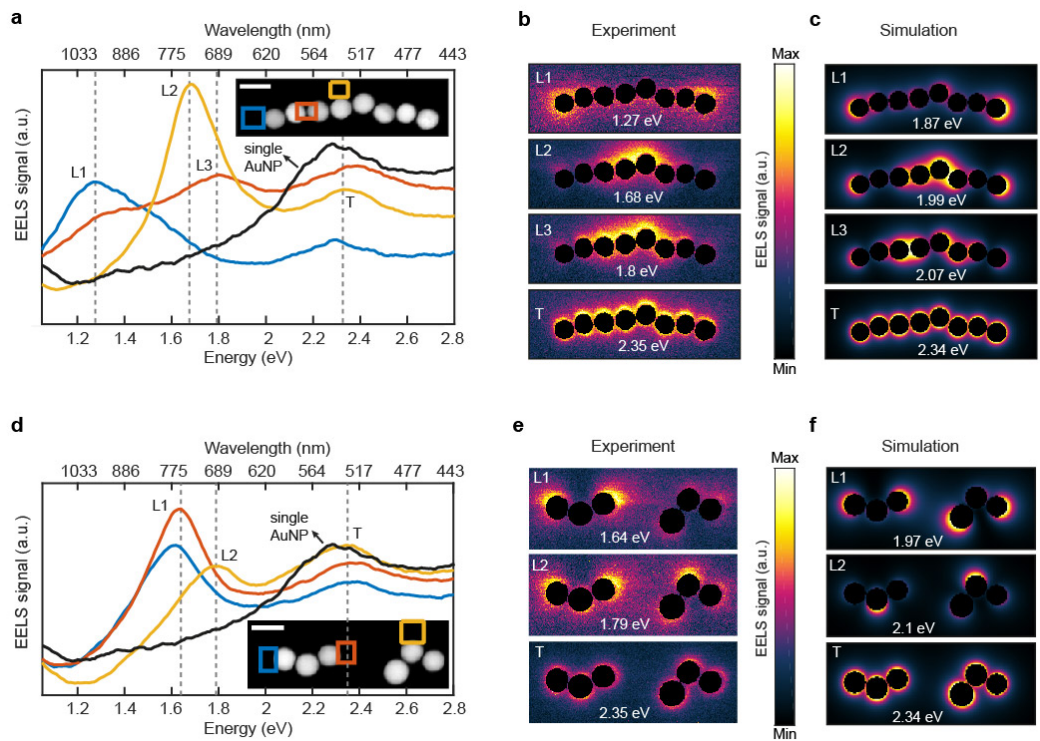

**Figure 2 | Electron energy loss characterization of complete and incomplete waveguides. (a)** Experimental EELS spectra of complete waveguide acquired within the squares in corresponding colours in the STEM micrographs insets. **(b, c)** Experimental and simulated EELS maps at different energies. **(d-f)** Same as **(a-c)** but for the incomplete waveguide. Propagating SP modes labelled according to their mode number (L: 1, 2, 3) from lowest to highest energy and longitudinal (L) / transverse (T) excitation in respect to the geometric arrangement. Scale bars, 50 nm.

The energy loss peaks for longitudinal (L) modes, where the induced dipoles are parallel to the nanoparticle chain, and transverse (T) mode, where dipoles are perpendicular to the AuNP chain[4] (Figure S 3), can clearly be distinguished. The energies of the L modes are red shifted while the energy of the T mode (2.35 eV, 527 nm) is slightly blue shifted compared to single AuNPs (2.3 eV, ~539 nm, see Figure S 4a) due to the strong SP coupling between AuNPs. The lowest energy mode L1, referred to as super-radiant mode[32,33], is detected at 1.27 eV (~976 nm). The higher energy L modes, L2 and L3, are detected at 1.68 eV (~738 nm) and 1.8 eV (~689 nm), respectively. L modes with even numbers (L: 2, 4, ...) are so-called dark modes or non-radiative modes as they do not have a net dipole moment, whereas bright modes (L: 1, 3, ...) have a net dipole moment[32]. Dark modes are also considered to effectively support propagation of SP modes along the closely spaced metal nanoparticle due to lower radiative losses than bright modes[33].

In Figure 2, we compare experimental EELS results (a, b, d, e) with boundary element method (BEM) simulations (c, f) that were carried out on the same particle arrangement as in the experiment (see Methods). The simulated modes are in good agreement with the experimental results. The nature of all modes can be clearly assigned (L: 1, 2, 3). The slight energy differences between experimental and simulation results (Figure 2) can be attributed to simplifications required for the simulation, *i.e.*, neglecting DNA capping on the AuNPs surface, assumption of perfect spheres, compensating for the silicon nitride substrate by an effective media or a different geometrical arrangement and interparticle spacing (Figure S 3). The occurrence of different modes resemble that from stochastically assembly AuNPs[32]. The incomplete waveguide reveals plasmon resonances of individual trimers whereas the complete waveguides support continuous SP propagation along the entire chain of eight particles (Supporting Table S 1).

We also examine complete and incomplete waveguides without a FND by CL imaging spectroscopy (setup see Figure S 5). The CL spectra show peak maxima at ~540 nm for the complete waveguide and at ~545 nm for incomplete waveguide (Figure

S 6). The CL spectrum is dominated by the transverse mode and is, as in the case of EELS, blue shifted compared to the single particle resonance energy (~557 nm, see Figure S 4d, simulated spectra in Figure S 7). Due to the low signal to noise ratio, the other modes cannot be identified clearly by CL.

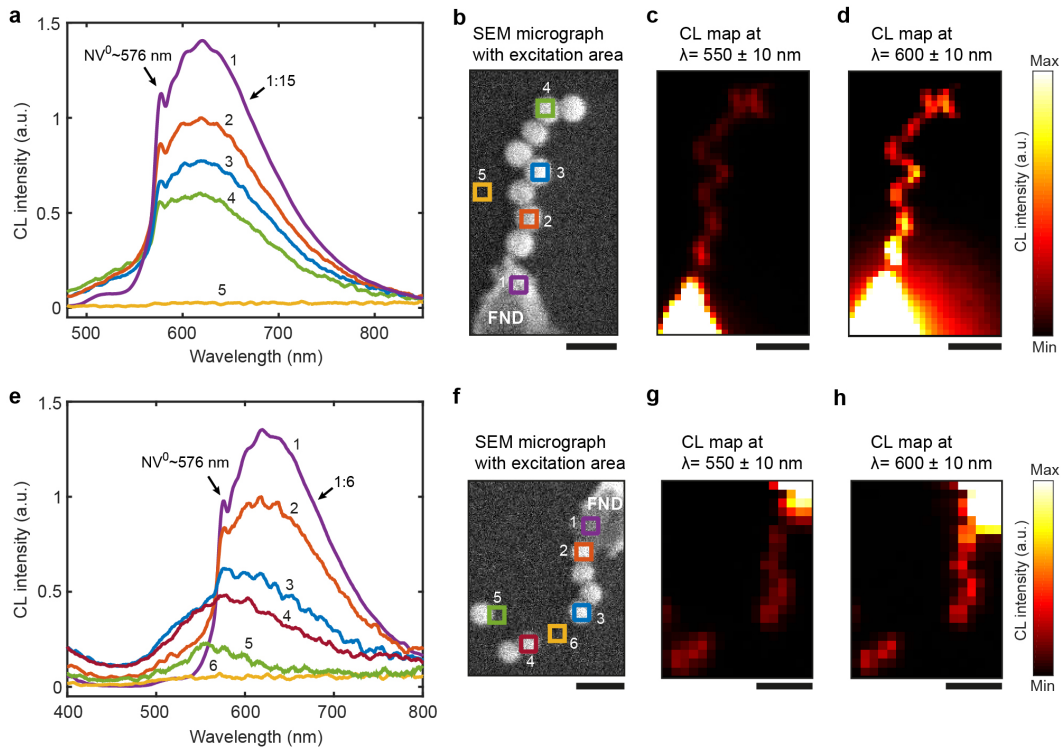

**Figure 3 | Energy transfer through plasmonic waveguides.** Cathodoluminescence (CL) spectra when illuminating waveguides adjacent to a fluorescent nanodiamond (FND). The CL spectra of the complete **(a)** and incomplete **(e)** waveguides taken at electron beam excitation areas indicated by squares in the corresponding colours in the SEM micrographs **(b, f)**. The intensity of the FND emission under direct e-beam illumination (purple) is reduced 15-fold **(a)** or 6-fold **(e)** for clarity. Spatially resolved CL maps at the detection wavelengths of 550 nm (± 10 nm) **(c, g)** and 600 nm (± 10 nm) **(d, h)**, showing CL intensity as a function of electron beam excitation positions. Scale bars, 100 nm.

Next, we demonstrate directed SP propagation through a waveguide towards a FND and detect it via excitation of FND emission, thus realising nanoscale frequency conversion. The FNDs are extremely photostable, bright, relatively insensitive to the chemical and physical environments and are therefore ideal for electron microscopy studies (see ref.[30,34]; and photoluminescence and CL spectra in Figure S 8). Moreover, they are promising nanostructures for single photon quantum information technology[30]. FNDs feature a nitrogen-vacancy (NV) point defect in the diamond lattice, where nitrogen replaces a carbon atom next to a vacancy, producing characteristic zero-phonon lines (ZPLs) at 575 nm for the neutral charge state ($NV^0$) and 637 nm for the negative charge state ($NV^-$).

Figure 3 shows the CL measurements on isolated complete and incomplete waveguides next to a single FND. In the complete waveguide, the $NV^0$ emission is observed when exciting any position in the waveguide, including the particle furthest from the FND (see also Figure S 9), which is approximately 350 nm away from the FND. Moreover, there is a significantly greater emission from the FND when the waveguide is directly illuminated, compared to illumination of the silicon substrate, as expected for an energy propagation though the particle chain.

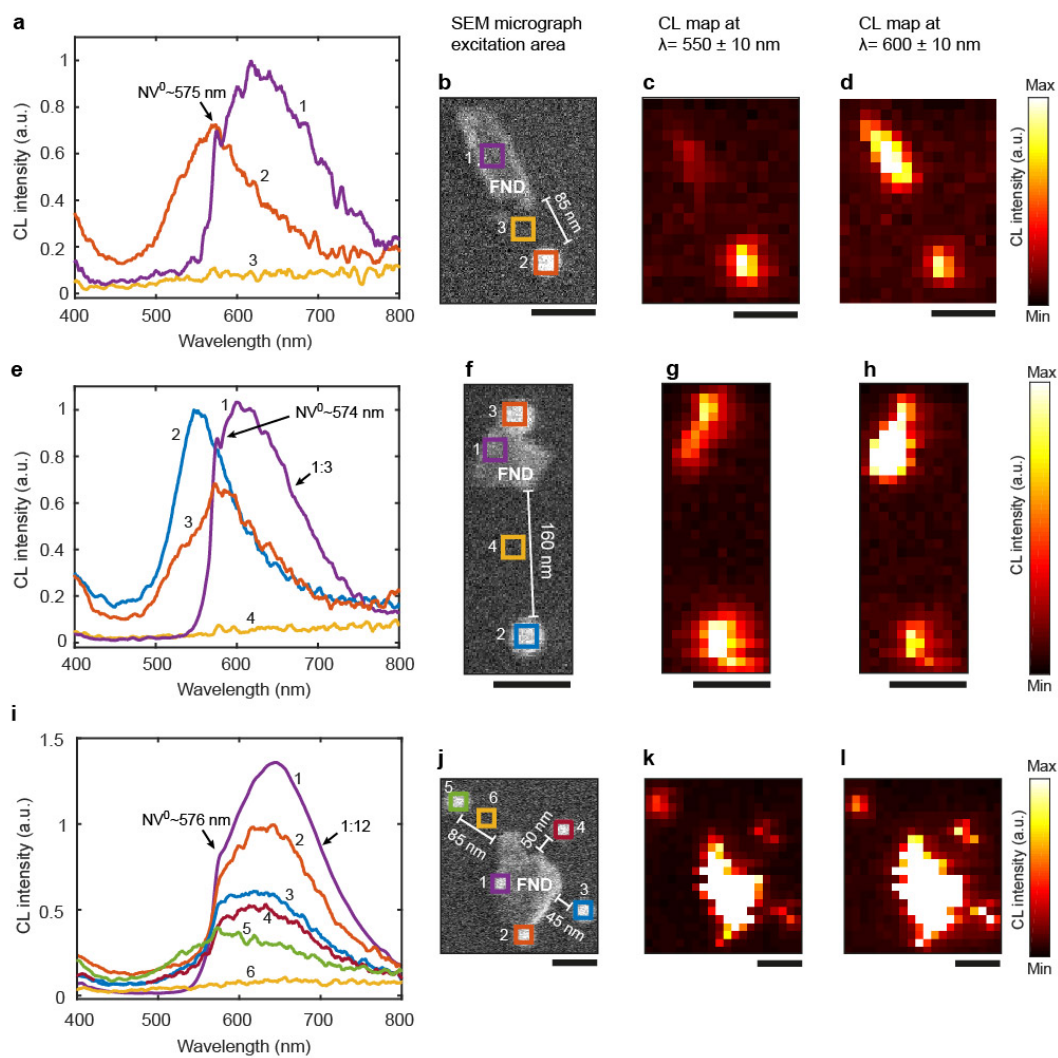

**Figure 4 | Cathodoluminescence measurements of single gold nanoparticles adjacent to a FND.** Cathodoluminescence (CL) spectra where the gold nanoparticle (AuNP) gaps are ~85 nm **(a)** and ~160 nm **(e),** and where five particles are scattered around a fluorescent nanodiamond (FND) with distances between 0-85 nm **(i)**. The spectra are taken at electron beam excitation locations indicated by squares in corresponding colours in the SEM micrographs **(b, f, j)**. The intensity of the FND spectra (purple) is reduced by 3-fold **(e)** and 12-fold **(i)** for clarity. Spatially resolved CL maps at the detection wavelength of 550 nm (± 10 nm) **(c, g, k)** and 600 nm (± 10 nm)

**(d, h, l),** showing CL intensity as a function of electron beam excitation positions (in pixels). Scale bars, 100 nm.

To confirm that indeed energy propagation through the AuNPs, and not an unrelated process such as a surface effect or scattering, is responsible for the emission of the FND, we analysed an incomplete waveguide as a negative control (Figure 3e). As expected, the particle closest to the FND excited the FND and exhibited a clear ZPL peak. Excitation of the particle before the gap (Figure 3e-f, spectrum 3) still showed a ZPL peak but the spectrum is broader and blue shifted compared to the FND emission and is likely a combination of the T mode and the FND emission. Moreover, the observed waveguide performance of the tetramer is lower than that of the octamer waveguide (see Figure 3), which can be explained by a lower coupling efficiency of the L3 and L4 mode to the $NV^0$ centres. The ZPL peak cannot be identified anymore in the CL spectrum when exciting the residual AuNP dimer beyond the ~82 nm gap (Figure 3e, spectrum 4). The broad spectrum shows a maximum at ~570 nm that is likely to be dominated by one of the longitudinal modes and the T mode. No FND emission can be detected for a single isolated AuNP (Figure 3e, spectrum 5), which was most likely detached from the incomplete 6-HB during sample preparation or drying. This AuNP is ~230 nm away from the FND, and its CL spectrum corresponds to that of an isolated particle (Figure S 4a). Further control experiments with isolated AuNPs in the vicinity of FNDs (Figure 4) confirm that no coupling to the FND takes place if an AuNP is ≥ ~85 nm away from the FND. However, shorter distances allow coupling to the FND to occur as indicated by a red shift compared to the CL spectrum of free AuNPs without a FND in their vicinity.

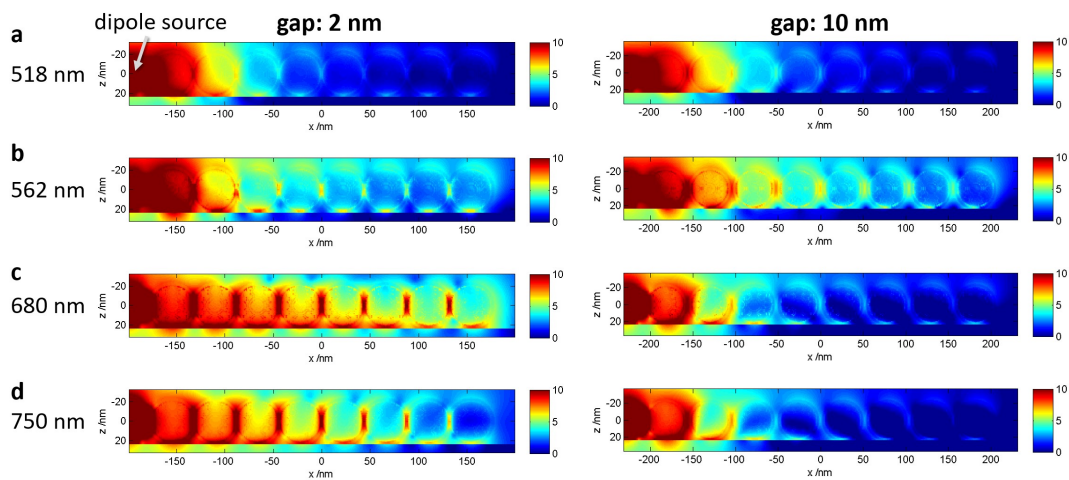

**Figure 5 | Energy propagation losses for different modes in a complete waveguide.** Integrated electric field and scaled by the integrated distance for **(a)** absorbing dominated mode at 518 nm (L4 mode), **(b)** scattering dominated mode at 562 nm (T mode), **(c)** scattering dominated mode 680 nm (L3 mode), and **(d)** scattering dominated mode at 750 nm (L2 mode). An interparticle gap of 2 nm (left) and 10 nm (right) was simulated. A long pulse length between 14 fs and 20 fs was selected for these mode calculations. Colour bars in logarithmic scale.

Initially, the basic T or L modes were thought to enable waveguiding[7,35], however, more recently, dark modes were also proposed as a mechanism.[33] We therefore simulated the losses of our specific system, taking into account the higher-order modes observed in EELS (Figure 5, Supplementary note 1). The electron beam was approximated by a static dipole source at one end of the waveguide. Our simulations reproduce the significant losses for the T mode (5.3 dB/50 nm at 562 nm) and L mode (L4: 5.3 dB/50 nm at 518 nm). However, losses of the lower energy modes were only 3.1 dB/50 nm at 680 nm (L3; Figure 5 c). We repeated the same simulations with an interparticle gap of 10 nm (Figure 5, right column). These scattering dominated and energetically lower SP modes (L2, L3) become very weak if the interparticle distances become higher than a few nm (Figure 5 right; Figure S 10), whereas the energetically higher modes (T, L4) are affected less and even tolerate removal of a particle (Figure S 11, Figure S 12). The results suggest that the L3-mode (73%) and L2-mode (10%) are dominating the energy transfer process. Since the excitation wavelength needs to be shorter than the emission of the $NV^0$ centre (shorter than 577 nm), coupling to the FND should be ineffective for the red-shifted modes. Distortion from the ideal, linear configurations, however, result in a blue shift (see Figure S 10), and the scattering cross section of the SP modes is so broad that we hypothesize they can still excite the $NV^0$ centre. Moreover, the interparticle gap of the experimental waveguides could be even smaller than 2 nm, as this cannot be detected from our electron micrographs.

In summary, we demonstrated that contrary to the common belief, nanoparticle chain waveguides can efficiently propagate energy over many hundred nanometers. We realized energy injection and subsequent energy propagation over a distance of 350 nm to a FND and outcoupling of energy to the far field, thus realizing a nanoscale light source and nanoscale energy conversion. . For this, we employed cathodoluminescence imaging spectroscopy that allowed for the study of the optical properties of the assemblies and energy propagation towards a fluorescent nanodiamond at a single-device level with nanometer resolution Our DNA-origami based self-assembly approach allowed the inter-particle spacing between highly monodispersed, single crystalline nanoparticles to be decreased down to 1-2 nm, which is not routinely achievable by lithography methods and is critical for efficient waveguiding. Electromagnetic modelling reveals that the energy transport along the particle chain is much more efficient *via* lower order longitudinal SP modes (L3 and L2 mode) than *via* the T or the super-radiant L1 mode and that the small inter-particle spacing is crucial. Whilst the centimeter scale propagation distances of silicon and dielectric waveguides remain unrivalled,[36] and metal stripes and wires can effectively guide plasmonic modes over many μm,[3] propagation lengths drop sharply when field confinement is below a few μm. A mode size of ~62 nm, which we demonstrate here, is not possible for strip waveguides[2,3,37] or silicon waveguides.

Finally, the precise and scalable self-assembly method presented herein can potentially be combined with an accurate positioning of DNA origami structures on silicon wafers[38]. We envision that this will enable the integration of nanometer-precise plasmonic components with larger and lower loss technologies for high-speed optical data transmission, quantum information technology, as single photon sources or sensing.

**Methods**

**Synthesis of spherical AuNPs.** The AuNPs were synthesised based on a seed-mediated growth and the protocol adopted from Steiner et al.[39] To form single-crystalline 2 nm Au seeds, 25 µL of 50 mM hydrogen tetra-chloroaurate (HAuCl$_4$, >99.9%, Sigma Aldrich) were added to 4.7 mL of 0.1 M hexadecyltrimethyl-ammonium bromide (CTAB, 99%, Merck KGaA) solution and slowly stirred for 10 min at 32 °C. Afterwards, 300 µL of freshly prepared 10 mM sodium borohydride (NaBH$_4$, 99%, Sigma Aldrich) were injected under vigorous stirring, which was continued for 30 seconds. For the next overgrowth step, the seeds were aged for 25 min at 32 °C to ensure the complete reaction of NaBH$_4$ and reproducibility.

To prepare 8 nm Au seeds, 20 mL of 200 mM of hexadecyltrimethylammonium chloride (CTAC, 25 wt.% in ultrapure water, Sigma Aldrich), 15 mL of 100 mM ascorbic acid (AA, C$_6$H$_8$O$_6$, >99%, Sigma Aldrich), and 1 mL of the initial CTAB-capped Au seeds were mixed in a 100 mL beaker. Subsequently, 20 mL of aqueous 0.5 mM HAuCl$_4$ solution was quickly injected under stirring. The reaction was allowed to continue for 15 min at room temperature (RT). The seeds were centrifuged at 15,000 rcf for 1 h and washed once more with ultrapure water. For the next round of growth, the Au seeds were dispersed in 10 mL of aqueous CTAC solution (10 mM). For this study, two batches of AuNPs were used with a diameter of 41.8 ± 2.6 nm or 44.9 ± 1.8 nm with similar results. To prepare CTAC-capped AuNPs, aqueous solutions of CTAC (60 mM, 200 mL), AA (1 M, 260 µL) and 2.5 mL (for a diameter of 41.8 nm), 1.95 mL (for a diameter of 44.9 nm) of the 8 nm seeds were mixed. Additionally, 200 mL of a growth solution including HAuCl$_4$ (1 mM) and CTAC (60 mM) were prepared and heated up to 45°C until the precursor complex was formed. Afterwards, the growth solution was added dropwise under moderate stirring using a syringe pump system at an injection rate of 1.0 mL/min. The reaction was allowed to continue for 12 h at RT after the injection was finished. The final product was collected by centrifugation at 12,000 rcf for 20 min and washed twice with 2 mM CTAC solution. For further use, the nanoparticles were re-dispersed in 30 mL of aqueous CTAC solution (5 mM). The respective size distribution (41.8 ±2.6 nm, 44.9±1.8 nm) was determined by TEM image analysis of at least 120 particles.

**Functionalization of AuNPs.** Before the functionalization of AuNPs with oligonucleotides, the CTAC cap was exchanged with bis (p-sulfonatophenyl)-phenylphosphine dihydrate dipotassium salt (BSPP, Sigma-Aldrich). For this, first 1 mL of the CTAC capped AuNPs was centrifuged at 10,000 rcf for 5 min to remove the excess CTAC, and they were dissolved in 1 mL of ultrapure water. After washing with ultrapure water, AuNPs were once more centrifuged at 10,000 rcf for 5 min, and resuspended in 1 mL of 5 mM BSPP supplemented with 1% Tween-20 (Applichem). The solution was then shaken for overnight. After that, 1 mL of methanol was added (the colour changed from red to blue), and centrifuged at 10,000 rcf for 5 min. The supernatant was discarded and AuNPs were resuspended in 1 mL of a 5 mM BSPP solution. After that, the AuNP solution was vortexed, and then shaken for 2-12 h at 25 rpm in a rotator. During this time, the solution was sonicated for 20 s several times until the AuNP solution turned red.

Next, AuNPs were conjugated to thiol-modified oligonucleotides (IDT), as described previously.[18] First, the thiol-modified oligonucleotides (5' ThioMC6-T15) were incubated with 20 mM TCEP (Tris (2-carboxyethyl) phosphine hydrochloride, Sigma-Aldrich) for 45 min. Then, AuNPs and the thiol-modified oligonucleotides were mixed in 0.5X TBE (Tris base, boric acid, EDTA, pH = 8.0) buffer at a ratio of 1:3000 (AuNP: ssDNA). To increase the oligonucleotide loading density on the surface of AuNPs, an aqueous 5 M solution of NaCl was added in four steps to reach a final concentration of 500 mM NaCl. After each addition of NaCl, the AuNPs solution was sonicated for 20 s and incubated for 20 min. After the last NaCl addition, the mixture of AuNPs and oligonucleotides was shaken for two days at 50 rpm with vibration mode at RT in a rotator. During this time, the solution was sonicated for a min several times to prevent unspecific dimer, trimer AuNP formation. After that, the solution was centrifuged at 10,000 rcf for 5 min to concentrate the AuNPs and supernatant was removed, and AuNPs were resuspended in 400 µL of 0.5X TBE buffer. Afterwards, the the excess of oligonucleotides was removed by ultrafiltration (100 kDa MWCO, Amicon Ultra, AMD Millipore) and washed five times with 400 µL of 0.5X TBE buffer at 10,000 rcf for 5 min right before mixing with 6-HBs.

**Folding of 6-HBs.** The same 6-HB design and staple strands were used as previously reported[5]. Briefly, to attach the thiol-modified poly-T conjugated AuNPs, 6-HBs were designed for eight binding sites with a centre-to-centre distance of 12 helical repeats (126 bp) resulting in 42.2 nm between two binding sites. Each binding site consisted of three single-stranded poly-A (5'-$A_{15}$) extensions of staple strands on three adjacent helices. For the incomplete waveguides, binding sites number 4 and 5 staple strands without poly-A extensions were used resulting in a centre-to-centre distance of 126.6 nm between binding sites number 3 and 6 (Figure S 1). To form the 6-HBs, 10 nM of p8064 scaffold (obtained from Dr. David Smith, IZI Leipzig), 168 short staple strands (IDT) (including respective binding sites as poly-A extensions) at 100 nM (each), 5 mM Tris (Applichem), 1 mM EDTA (Applichem) (pH 8) and 12 mM $MgCl_2$ (Applichem) were mixed. The mixture was annealed in a thermal cycler (Bio-Rad C1000 Touch) from 80 to 65 °C at a rate of -1 °C/min and from 65 °C to RT at a rate of -1 °C/20 min. After the folding reaction, 6-HBs were purified from excess staple strands by ultrafiltration (100 kDa MWCO) and washing five times with 400 µL of 1X TE containing 12 mM $MgCl_2$ at 10,000 rcf for 5 min.

**Waveguide assembly.** The purified 6-HB solution (0.8 nM final concentration) was quickly mixed with freshly functionalized and purified AuNPs (10 AuNPs/binding site) in 1X TE containing 12 mM $MgCl_2$ and incubated for 1 h at RT. Correctly assembled waveguides were separated from excess AuNPs and aggregates by agarose gel electrophoresis in a 0.75% agarose gel (Roche) containing 0.5X TBE buffer and 12 mM $MgCl_2$. The same buffer was used as the running buffer. All samples were mixed with a 20% volume of gel loading dye (50% glycerol, 5 mM Tris, 1 mM EDTA, 0.25% bromophenol blue and 0.25% of xylene cyanol) just before loading the samples to the gel pockets. Electrophoresis was performed in an ice bath at 4 °C for about 2 h at 70 V. The selected bands (Figure S 2) were cut out and chopped into small pieces. The waveguides were extracted from the gel with Freeze 'N Squeeze spin columns (Bio-Rad) by centrifugation at 5,000 rcf for 10 min.

To confirm the waveguide assembly, waveguide samples were imaged by SEM/tSEM. For this, carbon-coated TEM grids (carbon on formvar, Science Services) were plasma-treated for 20 s. Next, 5 μL of the waveguide sample solution was applied on the TEM grid and incubated for 5 min. The excess solution was removed from the grid with a filter paper. Next, 5 μL of a 2% uranyl formate solution was applied for a min to stain the 6-HB structures, and the solution was removed with a filter paper. The samples were scanned on Gemini SEM500 (Zeiss) and Helios 660 SEM/tSEM system (FEI) operated at 15 kV.

**EELS sample preparation.** Silicon nitride TEM membranes (10 nm thick, nine windows, SiMPore TEM Grids) were used as a thin substrate for EELS measurements. The TEM grids were plasma-treated for 3 min. Next, 10 μL of poly-L-ornithine solution (0.01 %, Sigma-Aldrich) was applied and incubated for 30 s and finally, the membrane was rinsed with ultrapure water and excess water was removed with a filter paper. Next, 10 μL of waveguide sample solution were applied and incubated for 2 min. Then the TEM membrane was washed once more with ultrapure water and the excess solution was removed with a filter paper.

**EELS measurements.** The EELS measurements were performed with a FEI Titan transmission electron microscope equipped with a monochromator and an image corrector. The microscope was operated in monochromated STEM mode at an acceleration voltage of 300 keV, providing a spot size of approximately 0.5 nm and an energy resolution of 0.10 eV (measured as the full-width at half-maximum of the zero-loss peak). The microscope was equipped with a Quantum 966 electron energy-loss spectrometer and the Gatan DigiScan acquisition system, which recorded an entire EELS intensity map in 5 to 25 min, depending on the number of pixels. A C3 aperture size of 50 μm, a camera length of 38 mm, an entrance aperture of 2.5 mm and a spectral dispersion of 0.01 eV per pixel was used. In addition, the automatic drift and dark current correction function included in the acquisition system was used. The individual EELS spectrum of the EELS intensity maps (with pixel sizes typically of 1–1.5 nm) were recorded with acquisition times ranging from 5 to 15 ms.

**EELS spectral processing:** To minimize the impact of monochromator drift during EELS acquisition, each row in the EELS data matrix was normalized to the EELS spectrum of the first column in the corresponding row. Afterwards, the zero-loss peak of each EELS spectrum was removed. To this end, we used two different methods, which both provide similar results: (i) the reflected-tail method, where the negative energy part of the zero-loss peak is mirrored around the zero-energy point to reconstruct the zero-loss peak, or (ii) fitting of a power-law function in the energy range 0.5 eV to around 1 eV to reconstruct the background signal. The resonant EELS intensity maps shown in this paper depict the summed background-removed EELS signal in a 0.15 eV spectral window centred at the resonance energies. The EELS signal from inside the nanoparticles was noisy and therefore removed in the map depictions. To detect the pixels inside the nanoparticles, we performed image analysis using the Image Processing Toolbox in MATLAB.

**EELS-BEM simulations.** Simulations of electron energy loss spectra and mappings were performed using the MATLAB MNPBEM13 toolbox[40], which is based on the boundary element method (BEM) open source code by F. J. Garcia de Abajo and A. Howie.[41] Each sphere of the waveguide was approximated by triangulation (400 vertices/particle). The dimensions and relative positions of the AuNPs were approximated to those observed in the TEM micrographs of the corresponding waveguide. The dielectric properties of gold were taken from P. B. Johnson and R. W. Christy.[42] We chose an effective medium (n = 1.3) to compensate the lack of the vacuum/substrate interface. The energy of the simulated electron beam was set to the experimental accelerating voltage of the TEM. For each waveguide, several spectra were evaluated in the energy range from 1-4 eV at different electron beam positions (non-penetrating) to ensure excitation of all possible plasmonic modes. Electron energy loss mappings were performed at selected energy levels and were simulated by a 2 nm mesh of the electron beam.

**CL sample preparation.** Silicon wafer pieces were used as substrates and were plasma treated for 10 min. Next, 20 µL poly-L-ornithine solution (0.01 %) were applied and incubated for 30 s. The substrate was then rinsed with ultrapure water. After removing the excess solution with a filter paper, 10 µL of waveguide sample solution was applied and incubated for 2 min. After that, the substrate was dried with a compressed air flow. For the waveguides with FNDs samples, after the waveguide deposition on the substrate, 10 µL of FND solution (FND in DI ultrapure water, high brightness 3.0 ppm NV, 0.1 wt%, 100 nm average particle size, Adámas Nanotechnologies) were applied and incubated for 2 min. Finally, the substrate was dried with a flow of compressed air.

**CL measurements and analysis.** The CL measurements were performed using a Tescan-SEM Delmic-CL setup. A focused 30 keV electron beam was scanned across the samples, which served as a nanoscale broadband source. The energetic electron beam induced an effective dipole that excited the nanostructures, with the resulting far-field emission collected by a parabolic mirror and directed into a spectrometer (Figure S 5). The electron beam was scanned across the sample with a ~15 nm pixel size, 1.12 nA current and a 10 µs dwell time. Exposure times ranged between 200 and 300 ms. To remove background CL signal originating from the silicon, reference measurements, taken from the exposed substrate, were subtracted.

The spectra were normalized, taking into account the spectral sensitivity of the system, and averaged over three pixels unless otherwise stated. Spectral filtering was carried out with a Savitzky-Golay function in MATLAB.

**FDTD simulations.** A commercial-grade simulator based on the finite-difference time-domain method was used to perform the calculations (Lumerical Inc., Version 8.16.1022)[43]. For the broadband and single mode excitation we used a dipole source for the specific wavelength range (short pulse length of 3 fs) and single wavelength (long pulse length between 14 fs and 20 fs), respectively. The electron beam is represented by a series of dipoles with a phase delay that is related to the electron velocity. We used an electron velocity with 20% of the speed of light and a dipole spacing of 15 nm. As boundary conditions, we used perfectly matched layers in all principal directions. Nanoparticle diameters, inter particle distance and materials constants were obtained from experimental studies. For the simulation ($\lambda = 400\ldots1000$ nm), the FDTD software approximates the refractive index indices of the materials by using a polynomial function. All optical constants were estimated by curve-fitting with a root mean square error below 0.21. For gold, we used the experimental data of Johnson and Christy[42]. As the substrate we used a silicon layer with refractive index data from Palik[44]. Zero-conformal-variant mesh refinement and an isotropic mesh overwrite region of 1 nm were used. All simulations reached the auto shut-off level of $10^{-5}$ before reaching 150 fs of simulation time.

**PL measurements.** For photoluminescence studies, the FND solution (Adámas ND-NV-100nm) was diluted in ultrapure water to a concentration of 0.2 mg/mL. 20 µL of the diluted FND solution were then drop cast onto the glass and silicon substrates. The PL measurements were performed on an inverted microscope equipped with a spectrometer (Ocean Optics, QE Pro). A 532 nm excitation wavelength from a Fianium super-continuum laser was chosen. The laser beam was focused onto a FND using an Olympus 100x 0.9NA microscope objective. The photoluminescence signal was then collected via the same objective. Several filters (Thorlabs DMLP 567 and FELH 550) were used in order to prevent any contribution from the excitation light to the collected signal.

**Supplementary Information** accompanies the paper

**Acknowledgements.** This work was funded by the DFG through the Center for Advancing Electronics Dresden (cfaed) as well as Seed Grant 043_2615A6 by the DFG Center for Regenerative Therapies Dresden (CRTD) to T.L.S.; by EPSRC (UK) (A.Z., C.P.T.McP., D.J.R.); the Royal Society and Wolfson Foundation. (S.R.); research grant (VKR023371) from VILLUM FONDEN (S.R.); a Multi University Research Initiative (MURIs FA9550-12-1-0488) from the AFOSR (S.R. and M.L.B.); the Volkswagen Foundation through a Freigeist Fellowship to T.K..; and the cfaed through an inspire grant (F.N.G.). We thank L. Eng, M. Mertig, A. Hille and A. Krasavin for helpful discussions; S. Gupta for producing artistic waveguides schemes; and the Biomod (biomod.org) team "Dresden DNAmic" (2014) for the help in the early stage of this project.

**The authors declare no competing financial interest.**

Correspondence and requests for materials should be addressed to T.L.S. (Thorsten-lars.schmidt@tu-dresden.de) or concerning simulations to T.K. (koenig@ipfdd.de).

# Supplementary information

## Supplementary note 1: Simulations of complete and incomplete waveguides excited by dipole source

In order to quantify the contribution of different SP modes to the waveguiding, we use finite-difference time-domain (FDTD) electromagnetic modelling. For this, we place a dipole source next to the complete and incomplete waveguides. An efficient excitation is observed for an electric field vector perpendicular to the surface of the first particle (see Figure S 13 for more details). Importantly, regardless of the source (dipole, e-beam, plane wave), the same waveguide modes are excited. We chose the dipole source here to study the waveguide performance and the nature of waveguide modes. An excitation with a broad dipole source with a broad spectrum (400 nm to 1000 nm wavelength) reveals four dominant modes for absorption and scattering caused by an energy loss at the band transition of gold (515 nm) and the radiant character of the energetic lower modes[33] (see Figure S 13). Single-mode excitation or long pulse length excitation of the dipole source, respectively, reveals the electric field at specific energies (see Figure 5 and Figure S 14). From these electric field calculations and the corresponding surface charge calculations we can determine the damping factor, which is the integrated electric field ratio between the first and last particle location and the nature of the SP modes respectively. The energetically highest mode (518 nm) can be identified as mode number L4, which is defined by the number of observed nodes ($n$-$1$). Such a node is defined by the sum of diminished net dipole moment and the first/last particle (see Figure S 15). Lower mode numbers (*L3* and *L2*) are observed for the scattering dominated modes at the wavelength 680 nm and 750 nm, respectively. The SP mode at 562 nm shows the signature of the transverse mode (T) because all net dipole moments point in the same direction. The proportion in scattering and absorption also corresponds to their damping factor, where *L4* mode (A: 518 nm), as well as transversal mode (B: 562 nm), shows a damping factor of 5.2 dB/50 nm, *L3* (C: 680 nm) 1.2 dB/50 nm and *L2* (D: 750 nm) 0.2 dB/50 nm (see Table S 1). Considering an incomplete waveguide (see Figure S 16, Figure S 11 and Figure S 12) the damping is increased for the 518 nm (A') mode (5.3 dB/50 nm), 562 nm (B') mode (5.2 dB/50 nm) and 686 nm (C') mode (see Table S 1). Importantly, the electric field significantly drops for the L3-mode, when we compare the electric field of the last particle between complete and incomplete waveguides.

| waveguide | complete | incompl. | complete | incompl. | complete | incompl. | complete | incompl. |
|---|---|---|---|---|---|---|---|---|
| wavelength | A: 518nm | A': 518nm | B: 562nm | B': 562nm | C: 680nm | C': 686nm | D: 750nm | D': n.a. |
| mode | L4 mode | n.a. | T mode | n.a. | L3 mode | | L2 mode | n.a. |
| x1=-150nm | 12259.30 | 12258.80 | 59415.40 | 59309.10 | 14007.40 | 16696.90 | 7238.43 | n.a. |
| x2=150nm | 1.94 | 1.88 | 41.50 | 35.83 | 190.73 | 14.91 | 26.40 | n.a. |
| ratio | 1.58E-04 | 1.54E-04 | 6.98E-04 | 6.04E-04 | 1.36E-02 | 8.93E-04 | 3.65E-03 | n.a. |
| contr. /% | 0.74 | 3.58 | 15.93 | 68.09 | 73.20 | 28.33 | 10.13 | n.a. |
| dB/300nm | -38.01 | -38.14 | -31.56 | -32.19 | -18.66 | -30.49 | -24.38 | n.a. |
| dB/50nm | -6.33 | -6.36 | -5.26 | -5.36 | -3.11 | -5.08 | -4.06 | n.a. |

**Table S 1 | Performance of complete and incomplete waveguides.** The damping factor (in dB unit) was determined by integrated electric field ratio (Figure S 5 and S 9) at first (x1=-150 nm) and last particle location (x2=-150 nm). The waveguide

performance was determined by the integrated electric field intensity at the last particle scaled by the sum of all intensities of the complete and incomplete waveguide, respectively.

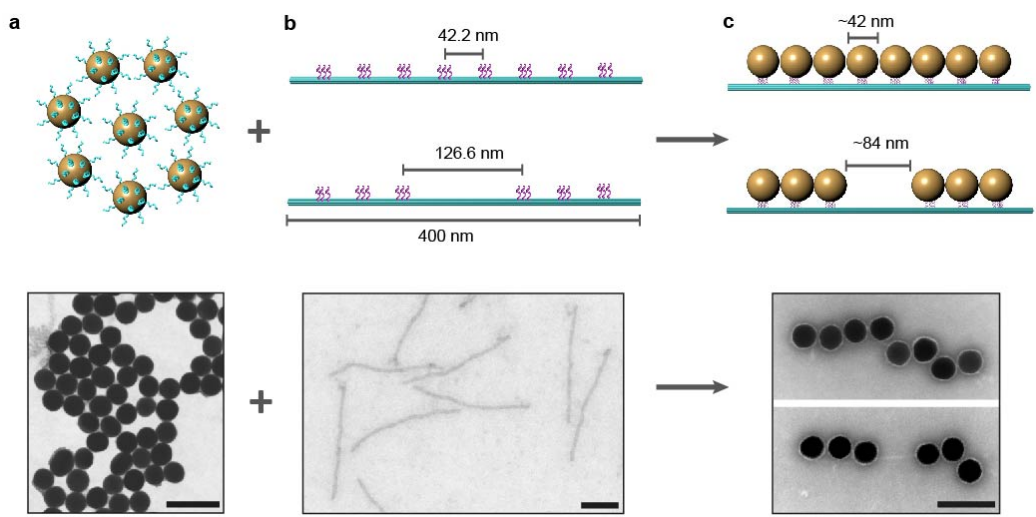

**Figure S 1 | Design and assembly reaction of waveguides: schemes and corresponding tSEM bright-field micrographs. (a)** AuNPs with a diameter of ~42 nm covered with thiol-modified oligonucleotides. (**b**) 6-HBs design with binding sites for complete (top) and incomplete waveguides (bottom). (**c**) Complete (top) and incomplete waveguide with a gap of ~84 nm (bottom). In the schematic representation the oligonucleotide shell around the AuNPs is omitted for clarity. Scale bars, 100nm.

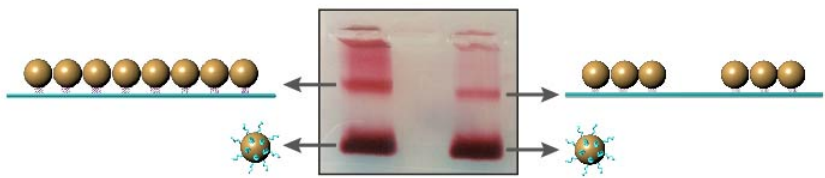

**Figure S 2 | Agarose gel electrophoresis purification of waveguides**. Real-colour image of agarose gel, showing the corresponding bands for the complete and incomplete waveguides, and excess AuNPs.

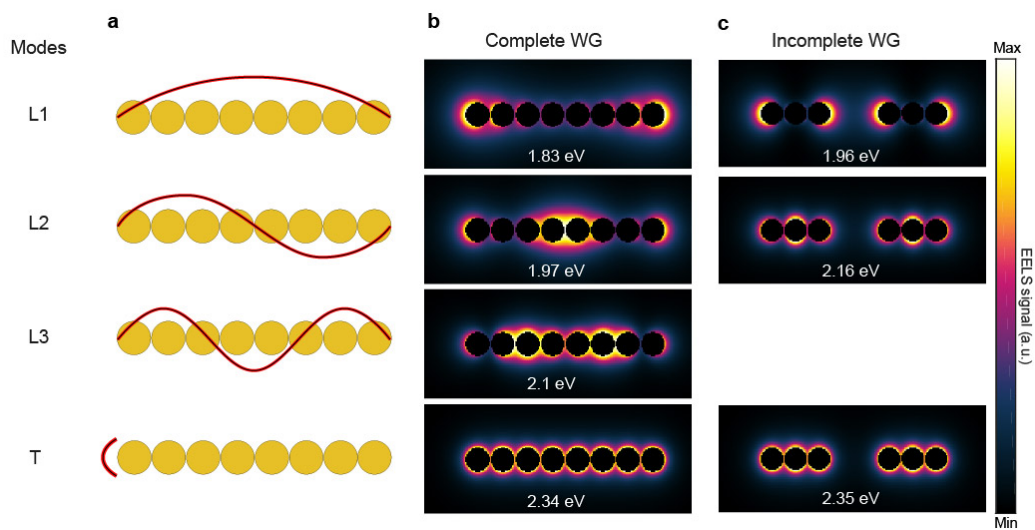

**Figure S 3 | Schematic of the SP modes of the complete waveguide (a). b-c)** EELS maps of linear complete and incomplete waveguides composed of spherical nanoparticles. Scheme illustrating the SP modes evolution in complete waveguide **(a)**. Simulated EELS maps **(b, c)** showing the longitudinal modes: super-radiant mode (L1), dark mode (L2), bright mode (L3), and transverse mode (T). The size of AuNPs is 42 nm with a 2 nm spacing. The gap between the trimer AuNPs is 90 nm.

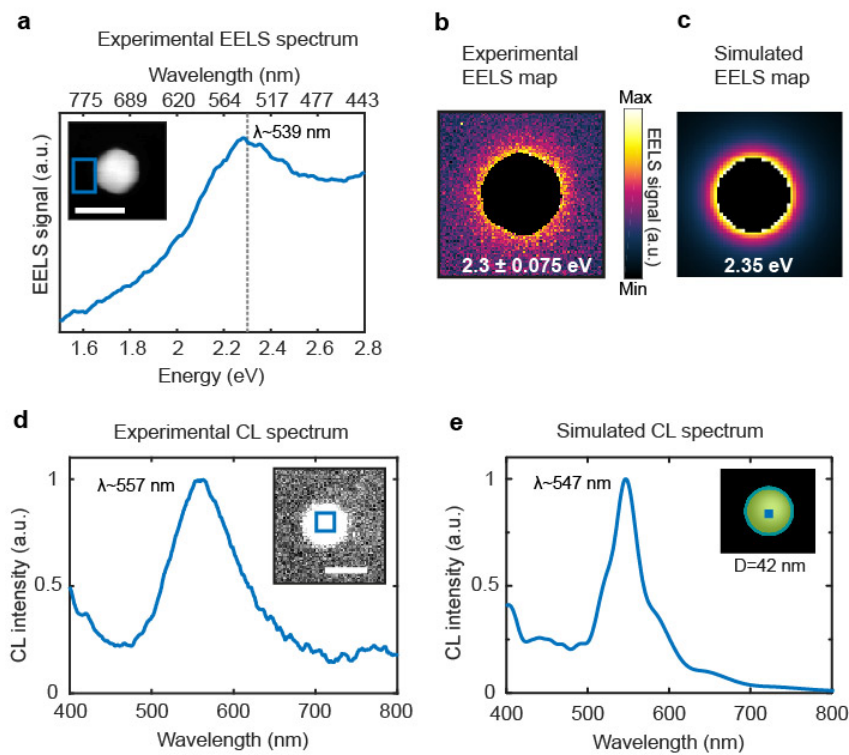

**Figure S 4 | Electron beam excitation of surface plasmons for ~42 nm single AuNP.** Experimental EEL spectrum of a single AuNP, taken in the location indicated by the blue square in the TEM micrograph (**a**). Experimental **(b)** and BEM simulation **(c)** EELS maps of the single AuNP at 2.3 eV and 2.35 eV, respectively. Experimental CL spectrum of a single AuNP, acquired for electron beam excitation location indicated by the blue square in the SEM micrograph inset **(d)**. FDTD CL simulation of a single AuNP, where the electron beam is located at the centre, as indicated by blue dot in modelled image **(e)**. The peak maxima are not identical due to the different substrates employed. Scale bars, 50 nm.

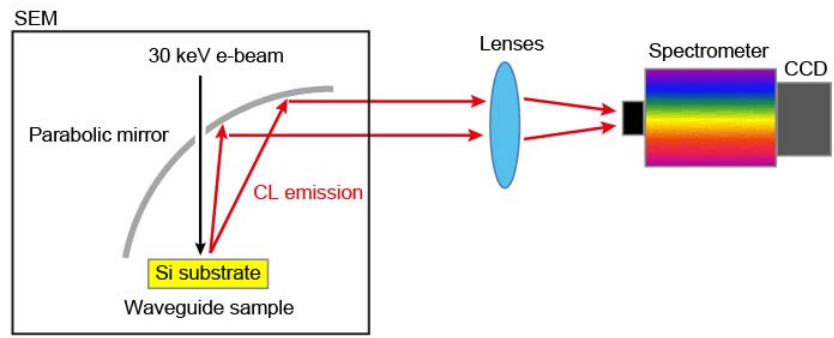

**Figure S 5 | Schematic of SEM-CL setup**. A 30 keV electron beam passes through a hole in a parabolic mirror and irradiates the sample, producing a cathodoluminescence signal. The photons emitted into the far-field are collected by a parabolic mirror and directed into the CL detector, which contains a spectrometer with a CCD camera. The electron beam is scanned across the sample, with the CL spectrum collected for each excitation position.

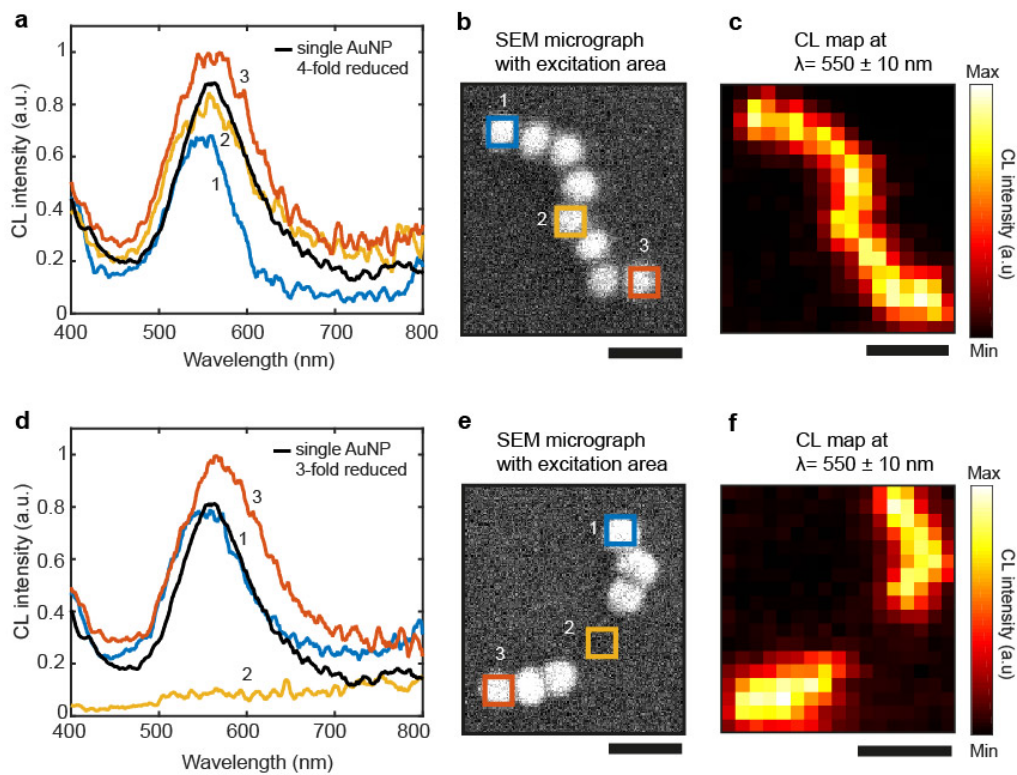

**Figure S 6| Cathodoluminescence imaging spectroscopy of complete and incomplete waveguides**. Experimental CL spectra **(a, d)** collected from the electron beam excitation locations indicated by the squares in the SEM micrographs **(b, e)**. Spatially-resolved CL maps of the waveguides at the detection wavelength of 550 nm ± 10 nm **(c, f)**, showing CL intensity as a function of electron beam excitation positions (pixel size ~15 x 15 nm). Scale bars, 100 nm.

Simulated CL spectra (Figure S 7) suggest that additional peaks may be seen in at wavelengths longer than 600-750 nm, but these are difficult to observe in the experimental spectra due to the low signal to noise ratio.

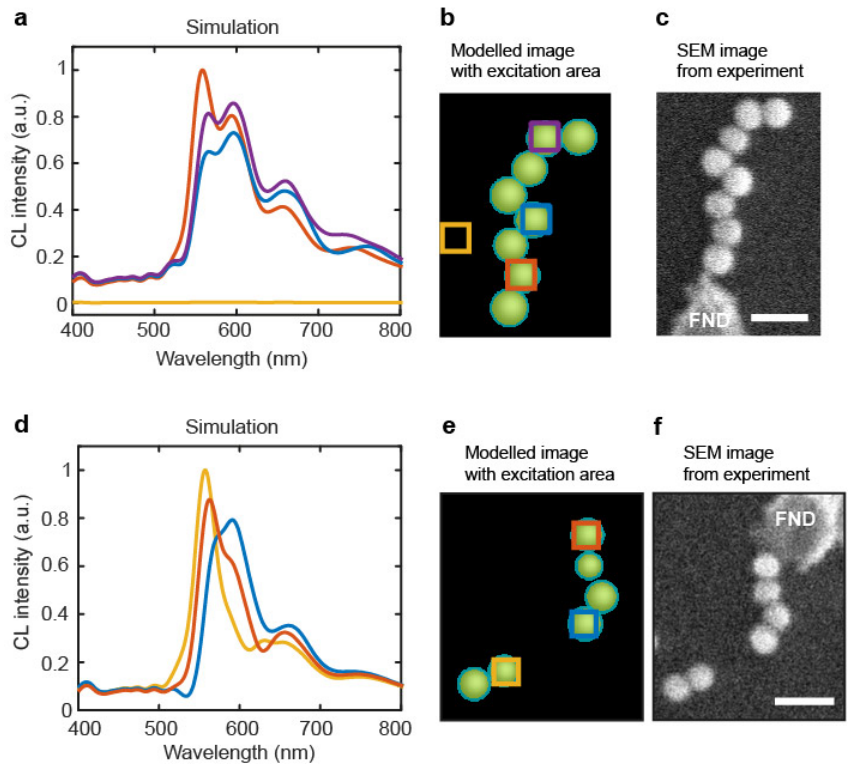

**Figure S 7 | FDTD simulations of normalized CL emission spectra of complete and incomplete waveguides without a FND**. Simulated CL spectra **(a, d)** are collected from the electron beam excitation locations are indicated by squares in corresponding colours in modelled images **(b, e)**. Simulated SEM micrographs **(c, f)** from experiments in Figure 3. Scale bars, 100 nm.

Differences between CL and EELS can be explained by the different substrates, specifically thin silicon nitride in the EELS measurements and silicon for CL, resulting in different effective refractive indexes.

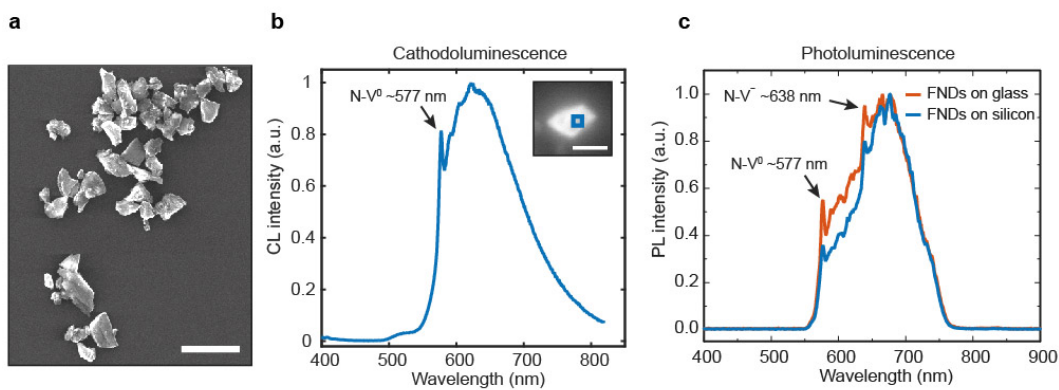

**Figure S 8 | Characterization of FNDs**. (**a**) SEM micrograph of the FNDs. Scale bar, 500 nm. (**b**) Normalized CL spectrum of a single FND with ZPL for $NV^0$ at 577 nm, inset SEM micrograph showing electron beam excitation location by the blue square. Scale bar, 100 nm. (**c**) PL spectra of FNDs solution on glass (red) and silicon substrate (blue) with ZPL lines for $NV^0$ at 577 nm and for $NV^-$ at 638 nm. The different excitation mechanisms give rise to different spectra, at the $N-V^-$ peak and lower energies.

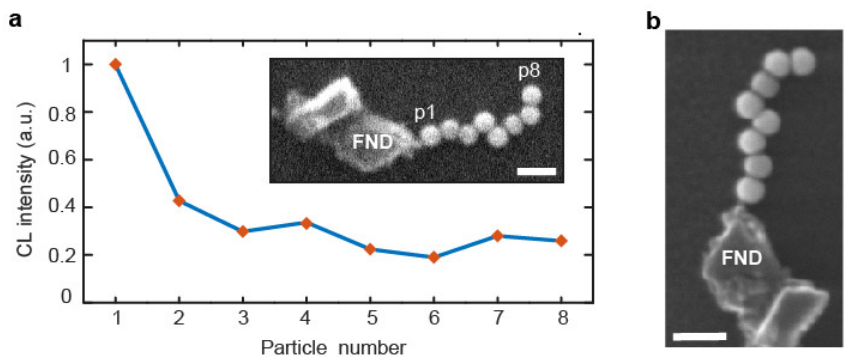

**Figure S 9 | CL intensity at the wavelength of 576 nm (corresponding to $NV^0$) as a function of excitation position in the complete waveguide. (a)** CL plot showing the ~ four-fold intensity decay of the emission over the length of the waveguide (~350 nm). (**b**) A high resolution SEM micrograph of the same waveguide with a FND, taken after the CL measurement, which confirms that the waveguide structure is still intact. The data are from Figure 3 a. Scale bars, 100 nm.

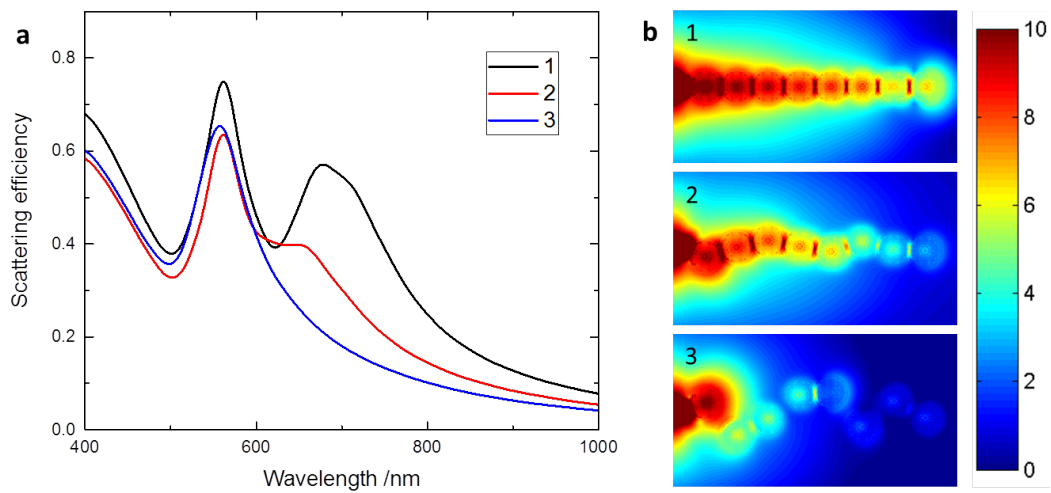

**Figure S 10 | Optical response induced by random placement of the particles.** (**a**) Scattering cross section efficiency for various random displacements of the particle positions. (**b**) Integrated electric field scaled the integrated area were calculated at 680 nm wavelength at the following random displacements: (**1**) perfectly aligned particle chain, (**2**) generated random displacement in *y*-direction with a defined standard deviation of 10 nm and (**3**) 20 nm. Colour bars in logarithmic scale.

The results indicate that small displacements result in a weakening and blue shift of the 680 nm mode and this mode collapses if particle distances become larger.

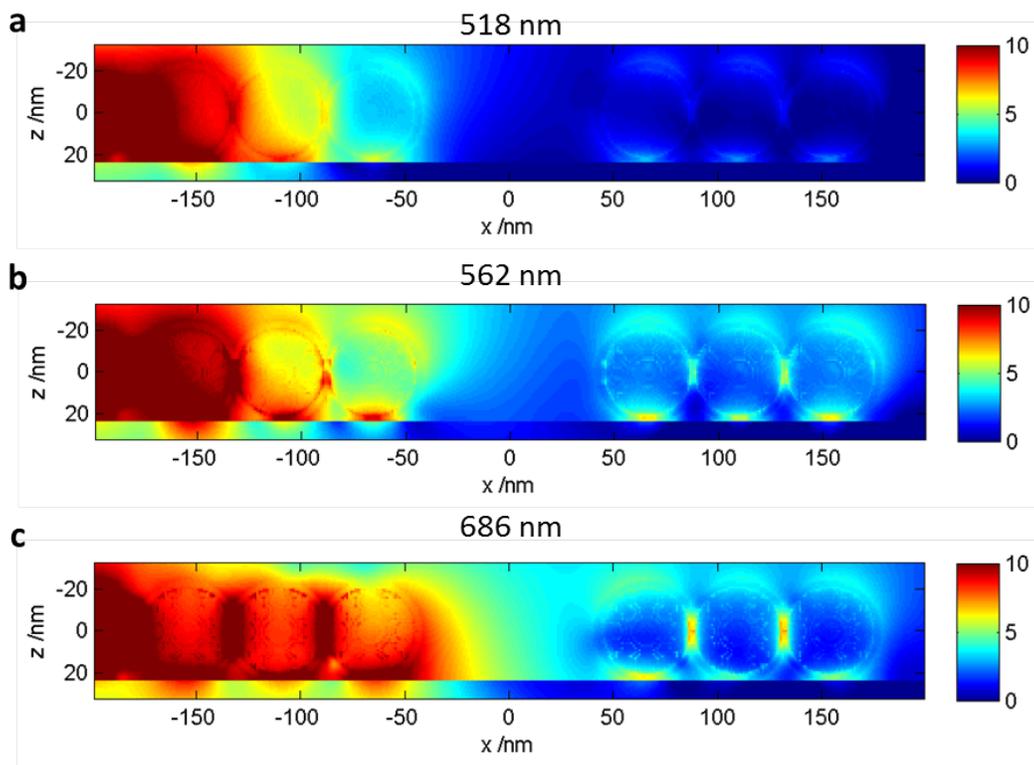

**Figure S 11 | Electric field images of selected SP modes for incomplete waveguide.** Integrated electric field for **(a)** absorbing dominated mode at 518 nm, **(b)** scattering dominated mode 562 nm, and **(c)** scattering dominated mode 686 nm. A long pulse length between 14 fs and 20 fs was selected for these mode calculations. Colour bars in logarithmic scale.

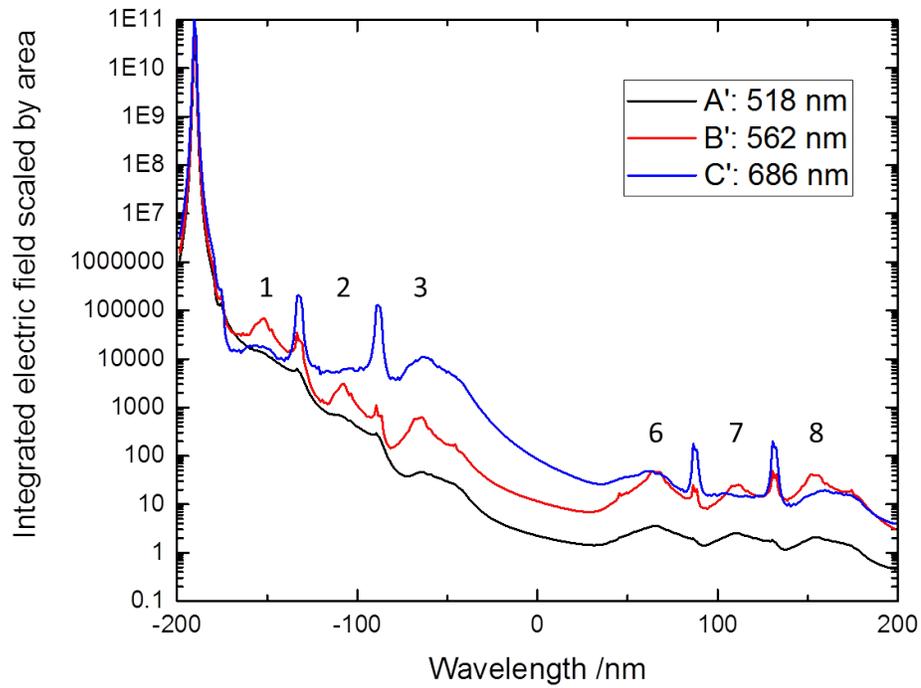

**Figure S 12 | Integrated electric field plot of various SP modes for incomplete waveguide.** Integrated electric field is scaled by the integrated area. The numbers between 1 and 8 represent the particle centre position. Particle 4 and 5 have been removed from simulations.

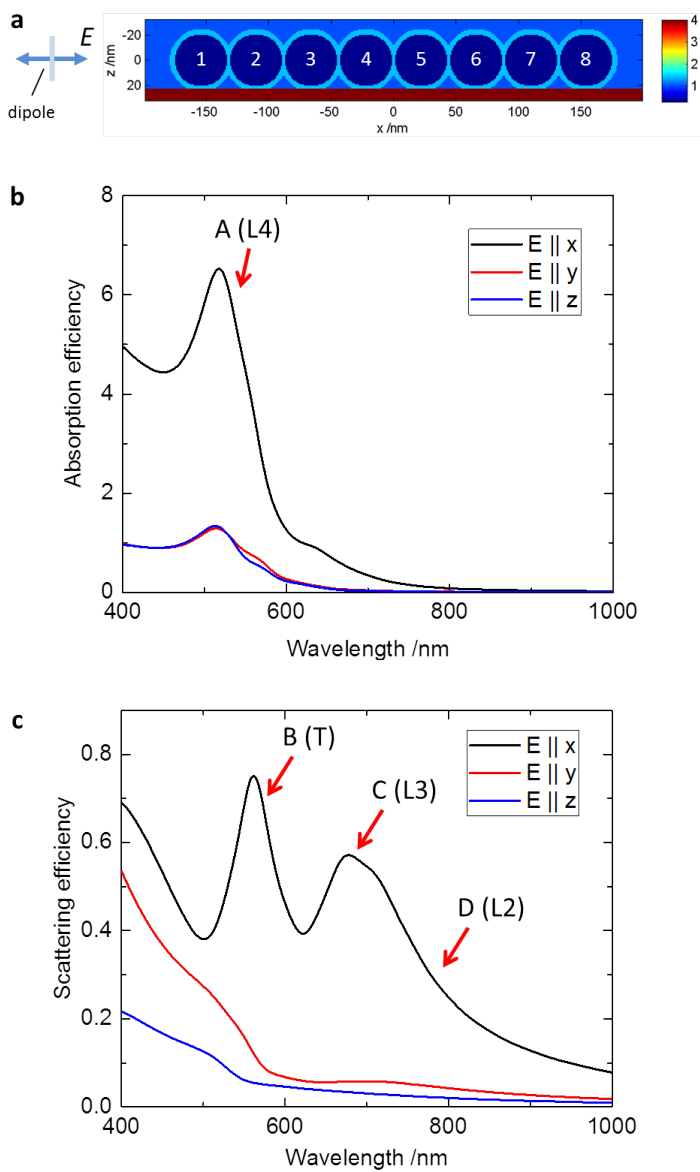

**Figure S 13 | Simulation of a complete waveguide excited by broad-band dipole source. (a)** Oriented dipole source with electric field vector and real refractive index cross section at 571 nm. Dark blue colour: gold spheres with 42 nm diameter and inter particle distance of 2 nm, light blue colour: 4 nm dielectric shell (refractive index of 1.4), blue colour: air, dark red: silicon substrate. The source is located at a distance of 15 nm from the first gold particle surface. **(b)** Absorption and **(c)** scattering cross section efficiency for various orientation of the dipolar electric field vector. The red arrow

highlights the dominant modes. A short pulse length of 3 fs was selected for this calculation.

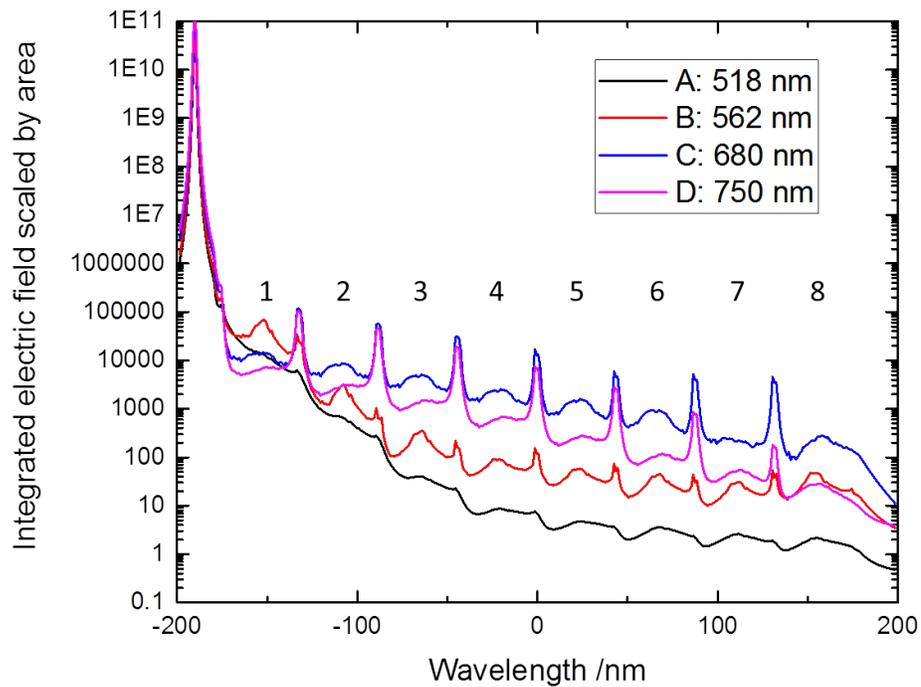

**Figure S 14 | Integrated electric field plot of various SP modes.** Integrated electric field and scaled by the integrated area. The numbers between 1 and 8 represent the particle centre position.

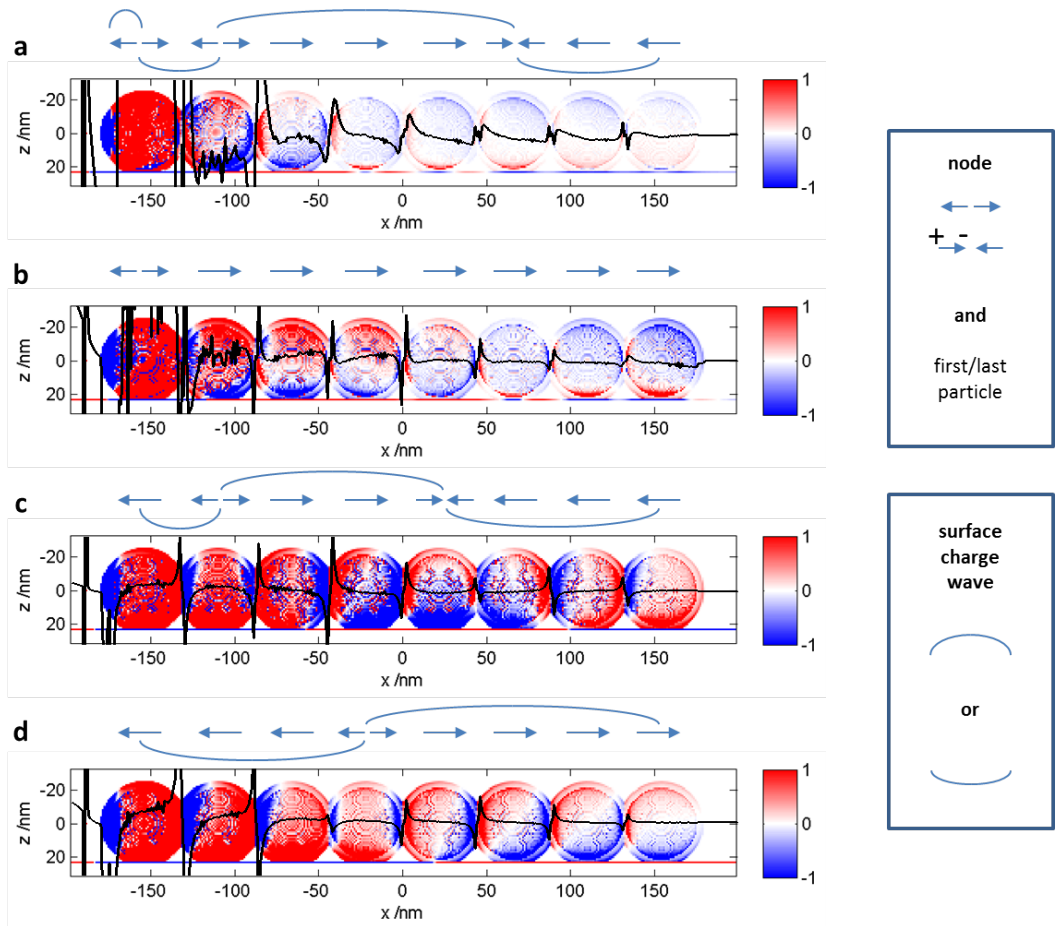

**Figure S 15 | Identification of SP modes by surface charges waves.** Integrated surface charge images and plots (black line) as well as corresponding net dipole moments and surface charge waves for **(a)** absorbing dominated mode at 518 nm (L4 mode), **(b)** scattering dominated mode at 562 nm (T mode), **(c)** scattering dominated mode 680 nm (L3 mode), and **(d)** scattering dominated mode at 750 nm (L2 mode). Red (+1) and blue (-1) colour indicates positive and negative charges, respectively.

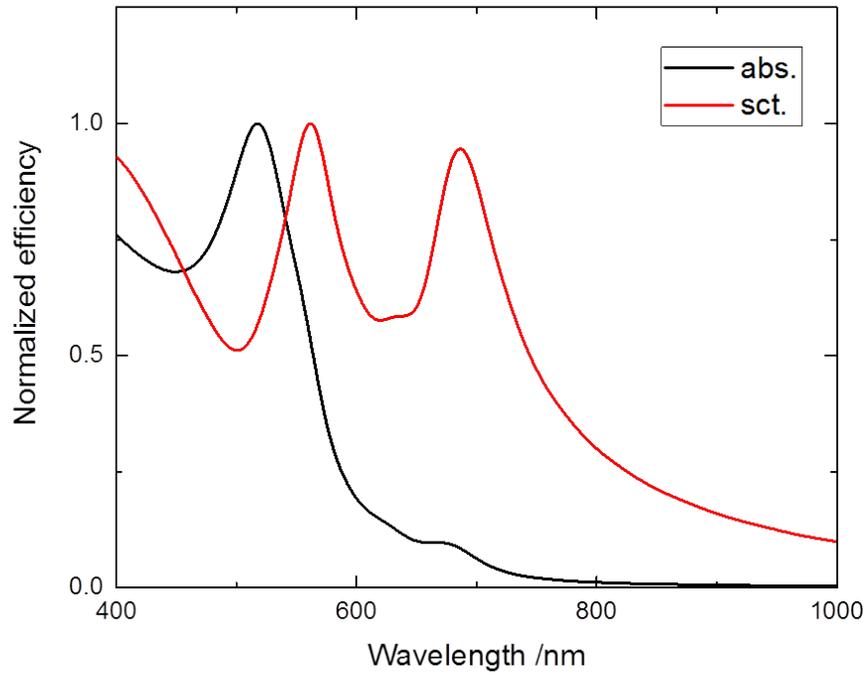

**Figure S 16 | Incomplete waveguide excited by broad-band dipole source.** Normalized absorption (black) and scattering (red) efficiency for an electric field vector orientation parallel to x-axis. The source is located at a distance of 15 nm from the first gold particle surface. A short pulse length of 3 fs was selected for this calculation.